\documentclass[12pt,letterpaper]{article}
\pdfoutput=1
\usepackage{jheppub}
\usepackage{amssymb}
\usepackage{color}
\usepackage{slashed}
\usepackage{cancel}
\usepackage{amsmath}
\usepackage{amsfonts}
\usepackage{amssymb}
\usepackage{graphicx,subcaption}
\usepackage{caption}
\usepackage{subcaption}
\usepackage{float}
\usepackage{tikz}
\usetikzlibrary[arrows]
\usepackage{amsmath}
\usepackage{epsfig}
\usepackage{appendix}
\usepackage{listings}
\usepackage{xcolor}
\usepackage{mathabx}

\def\bR{{\mathbb R}}

\def\bI{{\mathbb I}}

\newcommand{\be}{\begin{equation}}
\newcommand{\ee}{\end{equation}}
\newcommand{\vev}[1]{{\left< {#1} \right>}}

\title{A derivation of the planar limit of ${\cal N}=2$ chiral correlators}

\author{Bartomeu Fiol}
\author{and Alan Rios Fukelman}

\affiliation{Departament de F{\'\i}sica Qu\`antica i Astrof\'isica i \\Institut de Ci{\`e}ncies del Cosmos, 
Universitat de Barcelona,
Mart{\'\i}\ i Franqu{\`e}s 1, 08028 Barcelona, Catalonia, Spain}

\emailAdd{bfiol@ub.edu}
\emailAdd{ariosfukelman@icc.ub.edu}

\abstract{We derive analytically the terms of maximal transcendality of the planar 2- and 3-point functions of single-trace chiral primary operators of ${\cal N}=2$ SQCD on $\bR^4$, to all orders in the 't Hooft coupling. These results prove two conjectures we formulated in previous work. Furthermore, we also provide an explicit expression for the terms in the planar 2-point functions of these operators that contain products of two values of the $\zeta$ function.}

\begin{document}
\maketitle
\section{Introduction}

It is frustratingly difficult to compute analytically almost any quantity of interest in generic quantum field theories. The forecast is less gloomy for selected quantities in theories with extended symmetries, like supersymmetry and/or conformal invariance. In these cases, a variety of sophisticated techniques allows to explore analytically various questions, and in rare instances, to even obtain exact results.

In this regard, an arena that has proved to be particularly rich in recent years are four dimensional quantum field theories with ${\cal N}=2$ supersymmetry, and particularly those that simultaneously enjoy conformal invariance, ${\cal N}=2$ superconformal field theories. Out of the many quantities that one can consider in these theories, the so-called extremal correlators \cite{Papadodimas:2009eu} are particularly interesting. Indeed, symmetry completely fixes their space-time dependence, while it allows for very complicated dependence on the marginal couplings, which can be nevertheless determined in various regimes by complementary techniques. 

We will defer reviewing the precise definition of these correlators to the main body of the paper. For now, let's just recall that in the representation theory of the ${\cal N}=2$ superconformal algebra \cite{Dolan:2002zh} chiral primary operators (CPOs) are, by definition, those annihilated by all right chiral supercharges (similarly, anti-chiral operators are annihilated by all left chiral supercharges). On $\bR^4$, extremal correlators are the simplest non-trivial correlation functions of CPOs and anti-CPOs; they involve $n-1$ CPOs $O_i$ and a single anti-chiral operator $\bar O$ and take the form \cite{Papadodimas:2009eu}

\be
\vev{O_{\Delta_1}(x_1) \dots O_{\Delta_{n-1}}(x_{n-1}) \bar O_{\bar \Delta}(y)} = \frac{\vev{O_{\Delta_1} \dots O_{\Delta_{n-1}} \bar O_{\bar \Delta}} (\tau,\bar \tau)}{|x_1-y|^{2\Delta_1}\dots |x_{n-1}-y|^{2\Delta_{n-1}}}
\ee 
with $\Delta_1+\dots \Delta_{n-1}=\bar \Delta$. Note that the space-time dependence of these correlators is completely fixed. All that remains to be determined are the coefficients $\vev{O_{\Delta_1} \dots O_{\Delta_{n-1}} \bar O_{\bar \Delta}} (\tau,\bar \tau)$ which are functions of the marginal couplings. In this work, we will consider Lagrangian ${\cal N}=2$ SCFTs with a single gauge group, and then $\tau$ is the complexified gauge coupling $\tau=\frac{2\theta}{\pi}+i\frac{4\pi}{g_{\text{YM}}^2}$.

One of the tools available to determine these coefficients is supersymmetric localization \cite{Pestun:2007rz} (see \cite{Pestun:2016zxk} for a review). More specifically, it was argued in \cite{Gerchkovitz:2014gta, Gerchkovitz:2016gxx} that supersymmetric localization reduces the evaluation of closely related n-point functions on $S^4$ to matrix model computations; then, a Gram-Schmidt orthogonalization procedure applied to these $S^4$ correlators yields the correlators on $\bR^4$ \cite{Rodriguez-Gomez:2016ijh, Rodriguez-Gomez:2016cem, Billo:2017glv, Beccaria:2020hgy,Galvagno:2020cgq, Beccaria:2021hvt}. Other approaches use the 4d analog \cite{Papadodimas:2009eu} of the $tt^*$ equations \cite{Baggio:2014sna, Baggio:2014ioa, Baggio:2015vxa, Baggio:2016skg}, the large R-charge expansion \cite{Hellerman:2015nra, Hellerman:2017sur, Bourget:2018obm, Grassi:2019txd, Beccaria:2018xxl, Beccaria:2020azj, Hellerman:2021yqz} or holography \cite{Billo:2021rdb, Billo:2022fnb}. 

Our approach to compute these extremal correlators - initiated in our previous work \cite{Fiol:2021icm} - also relies squarely on supersymmetric localization, but it contains a couple of twists that we now detail. The first new ingredient is an alternative approach to compute the relevant integrals \cite{Billo:2017glv,Billo:2018oog,Billo:2019fbi} - dubbed the full Lie algebra approach in \cite{Fiol:2018yuc} - that doesn't restrict the integration domain to a Cartan subalgebra, thus avoiding the introduction of Vandermonde determinants. For ${\cal N}=2$ theories, this leads to a rewriting of the matrix integrals in terms of an effective action with an infinite number of single and double trace terms \cite{Billo:2019fbi, Fiol:2020bhf}. The second ingredient is a combinatorial solution of the planar limit of such matrix models with single and double trace terms \cite{Fiol:2020bhf, Fiol:2020ojn,Fiol:2021icm,Fiol:2021jsc}, given by a sum over tree graphs. 

In our previous work \cite{Fiol:2021icm}, we applied the approach just summarized to planar 2- and 3-point functions of single-trace operators of ${\cal N}=2$ SQCD, {\it i.e.} SU(N) super Yang-Mills with $\textnormal{N}_F=2 \textnormal{N}$ massless hypermultiplets in the fundamental representation. We were able to derive the planar limit of these 2- and 3- point functions on $S^4$. On the other hand, a straightforward application of the Gram-Schmidt orthogonalization procedure to derive the equivalent correlators on $\bR^4$ becomes cumbersome quite rapidly, as one increases the conformal dimension of the operators under study. To understand what was accomplished in \cite{Fiol:2021icm}, it is relevant to bear in mind that the supersymmetric localization approach reveals that in the planar limit these coefficients take the form
\be
\vev{O_k \widebar{O_k}} = k \tilde \lambda^k \left[1+\sum_{m=1}^\infty \sum_{n_1,\dots,n_m=2}^\infty a_k(n_1,\dots,n_m) \zeta_{2n_1-1}\dots \zeta_{2n_m-1} \tilde \lambda^{n_1+\dots+n_m} \right]
\label{generictwo}
\ee

\be
\frac{\vev{O_{k_1} O_{k_2} \bar{O}_{k_1+k_2}}_n}{\sqrt{k_1 \cdot k_2 \cdot (k_1+k_2)}} 
=\frac{1}{\textnormal{N}} \left[ 1+\sum_{m=1}^\infty \sum_{n_1,\dots,n_m=2}^\infty b_{k_1,k_2}(n_1,\dots,n_m) \zeta_{2n_1-1}\dots \zeta_{2n_m-1}  \tilde \lambda^{n_1+\dots+n_m} \right]
\label{genericthree}
\ee
with $\zeta_i$ values of the $\zeta$ function, $\tilde \lambda=\frac{g_{\text{YM}}^2 \textnormal{N}}{16\pi^2}$ the reduced 't Hooft coupling and $a_k(n_1,\dots ,n_m)$ and $b_{k_1,k_2}(n_1,\dots,n_m)$ rational numbers. In \cite{Fiol:2021icm} we computed by brute force all the terms with a single value of $\zeta$ - see ({\ref{generictwo}), (\ref{genericthree}) - for $\langle O_2 \bar O_2\rangle$,  $\langle O_4 \bar O_4\rangle$, $\langle O_6 \bar O_6\rangle$ and $\langle O_2 O_2 \bar O_4 \rangle_n, \langle O_2 O_4 \bar O_6 \rangle_n$. The results turned out to be strikingly simple, and motivated two conjectures for the coefficients $a_k(n)$, and $b_{k_1,k_2}(n)$ in ({\ref{generictwo}), (\ref{genericthree}), that we recall below.

In this work we take a more systematic approach to the Gram-Schimdt orthogonalization procedure, and this yields two main results. First we derive all terms with just a single value of $\zeta$ in (\ref{generictwo}), (\ref{genericthree}), thus proving the conjectures of \cite{Fiol:2021icm}.  Moreover, for the 2-point functions, we derive a reasonably explicit formula for terms with two values of $\zeta$ in (\ref{generictwo}). Specifically,

\be
\begin{split}
&\langle O_{2k}\bar O_{2k} \rangle = 2k \tilde \lambda^{2k} \left[1-4k\sum_{n=2}^\infty \frac{\zeta_{2n-1} (-\tilde \lambda)^n}{n}
{2n \choose n} \left[ {2n \choose n+2k}+{2n \choose n+1}-n \right]  \right.\\
& +4k \sum_{m,n=2}^\infty \frac{\zeta_{2m-1} \zeta_{2n-1}}{mn}(-\tilde \lambda)^{m+n} {2m \choose m}{2n\choose n} \bigg(  m\, n \bigg[-({\cal C}_m-1)({\cal C}_n-1)  \\
& \left. +({\cal C}_m-1) \left( {2n \choose n+2k}+{2n-1 \choose n}-n\right)+
 ({\cal C}_n-1) \left( {2m \choose m+2k}+{2m-1 \choose m}-m\right)\right] \\
& +2k\left[{2m \choose m+2k}+m({\cal C}_m-1)\right]\left[{2n \choose n+2k}+n({\cal C}_n-1)\right] \\
& +2n^2 \sum_{i=1}^{n-1}{n-1 \choose i}{n-1 \choose i-1} \sum_{j=1}^{m-1}{m\choose j}{m \choose j-1} \frac{1}{i+j} \\
& \left. \left. +2k \sum_{i=1}^{n-1}{n \choose i+k}{n \choose i-k} \sum_{j=1}^{m-1}{m\choose j+k}{m \choose j-k} \frac{k^2+ij}{i+j} \right)+\dots \right]
\end{split}
\label{2result}
\ee

and 

\be
\begin{split}
&\langle O_{2k_1}O_{2k_2}\bar O_{2k_3}\rangle_n = (2k_1 2k_2 2k_3)^{1/2}\\
&\left [1-\sum_{n=2}^\infty \zeta_{2n-1} (-\tilde \lambda)^n {2n \choose n}
\left ({2n \choose n+2k_1}+{2n \choose n+2k_2}+{2n \choose n+2k_3} +(n-1)\left({\cal C}_n-2\right)\right ) +\dots \right]
\end{split}
\label{3result}
\ee
where ${\cal C}_n$ are Catalan numbers. The terms with a single $\zeta$ in (\ref{2result}) and (\ref{3result}) were already conjectured in \cite{Fiol:2021icm}. As it happens in similar computations, (\ref{2result}) illustrates that terms with maximal trascendality ({\it i.e.} with a single value of $\zeta$) are much simpler than generic terms.

This work leaves open a number of questions. It might be worth to extend the analysis presented here for ${\cal N}=2$ SCQD to extremal correlators of other Lagrangian ${\cal N}=2$ SCFTs \cite{Fiol:2015mrp, Beccaria:2020hgy, Galvagno:2020cgq, Beccaria:2021hvt}. While in this paper we have managed to prove the conjectures in \cite{Fiol:2021icm}, we still think that it should be possible to derive these results directly on $\bR^4$, perhaps diagrammatically; see \cite{Mitev:2014yba, Beccaria:2020hgy, Galvagno:2020cgq} for analysis of the potentially relevant diagrams.

The structure of the paper is the following. In section 2 we recall the relation between $S^4$ and $\bR^4$ extremal correlators. In section 3 we consider planar two-point functions of extremal correlators; we derive explicit formulas for terms with one or two values of $\zeta$. In section 4 we derive the analytic form of the terms with a single value of $\zeta$ in the planar 3-point function. Two appendices contain various technical details.

\section{From $S^4$ to $\bR^4$ chiral correlators}
The main goal of this paper is to compute correlators of certain chiral operators of ${\cal N}=2$ SU(N) SQCD on $\bR^4$. Let's start recalling what these operators are, and spelling out the strategy to compute their correlation functions.

The ${\cal N}=2$ vector multiplet contains a complex scalar $\phi$, and one can consider single trace operators of the form $O_m  \, \propto \, \text{Tr } \phi^m$. These operators are chiral primary operators (CPOs), {\it i.e. } they are annihilated by all the right chiral supercharges. Thus, their conformal dimension is fixed by their $u(1)_R$ R-charge, and in this case they have $\Delta=m$. One can analogously define anti-chiral primary operators \cite{Papadodimas:2009eu, Dolan:2002zh}.

Symmetry severely constrains the form of n-point functions of CPOs and anti-CPOs. For instance, on $\bR^4$, these n-point functions vanish unless the total R-charge is zero. Thus, the simplest non-trivial n-point functions are the so-called extremal correlators, corresponding to the case of $n-1$ CPOs $O_i$ and one anti-CPO $\bar O$. The spacetime dependence of extremal correlators is completely fixed,
\be
\vev{O_{\Delta_1}(x_1) \dots O_{\Delta_{n-1}}(x_{n-1}) \bar O_{\bar \Delta}(y)} = \frac{\vev{O_{\Delta_1} \dots O_{\Delta_{n-1}} \bar O_{\bar \Delta}} (\tau,\bar \tau)}{|x_1-y|^{2\Delta_1}\dots |x_{n-1}-y|^{2\Delta_{n-1}}}
\ee 
with $\Delta_1+\dots \Delta_{n-1}=\bar \Delta$ and one is left with the task of computing the coupling-dependent coefficients.

Given the power of supersymmetric localization to simplify the computation of selected quantities in ${\cal N}=2$ theories, it is natural to ask  whether it can be applied to the determination of these coefficients. This was answered in the positive in 
\cite{Gerchkovitz:2016gxx}, who showed that for a suitable deformation of the original SCFT, placed on $S^4$, supersymmetric localization allowed to reduce the computation of chiral correlators on $S^4$ to a zero-dimensional problem. However, due to operator mixing, one can't just identify the operators on $S^4$ with the ones we have introduced in $\bR^4$. To undo this mixing, \cite{Gerchkovitz:2016gxx} proposed to apply a Gram-Schmidt orthogonalization procedure. Specifically, and focusing on operators of even dimension, denote by $\Omega_{2\ell}$ the operators on $S^4$ and by $O_{2k}$ the chiral operators on $\bR^4$. Next, define the Gram-Schmidt determinant
\be
D_k = 
\begin{vmatrix}
\langle \Omega_2 \Omega_2 \rangle & \dots & \langle \Omega_2 \Omega_{2k} \rangle \\
\dots & \dots & \dots \\
\langle \Omega_{2k} \Omega_2 \rangle & \dots & \langle \Omega_{2k} \Omega_{2k} \rangle
 \end{vmatrix}
\label{gsdet} 
 \ee
where the entries are two-point functions of operators on $S^4$. Then the orthogonal basis of operators on $\bR^4$ can be formally written in terms of determinants where the last row is given by $\Omega_{2\ell}$ operators 
 \be
 O_{2k}= \frac{1}{D_{k-1}}
 \begin{vmatrix}
 \langle \Omega_2 \Omega_2 \rangle & \dots & \langle \Omega_{2k} \Omega_{2} \rangle \\
\dots & \dots & \dots \\
\langle \Omega_2 \Omega_{2(k-1)} \rangle & \dots & \langle \Omega_{2k} \Omega_{2(k-1)} \rangle \\
\Omega_2 & \dots & \Omega_{2k}
\end{vmatrix}
\label{def:o2k}
\ee
The strategy is now clear. First, use supersymmetric localization to compute two-and three-point functions on $S^4$, and then apply the Gram-Schmidt procedure to deduce from these the correlators on $\bR^4$. In practice, the first step is extremely hard to implement since the relevant integrals involve both perturbative and non-perturbative contributions. As for the second step, while it is easy to implement for operators with small conformal dimension, it gets out of hand as we consider operators with increasing conformal dimension. In our previous paper \cite{Fiol:2021icm} we solved the first issue in certain regime, and in the present paper we will make some progress with the second one.

In \cite{Fiol:2021icm} we restricted to the planar limit of the gauge theory, and we derived the planar limit of the 2- and 3- point functions on $S^4$ as all-order perturbative series in the 't Hooft coupling $\lambda$. To display them, first define $\tilde \lambda =\frac{\lambda}{16 \pi^2}$,
\be
f(i)=\frac{(2i)!}{i! (i-1)!} \tilde \lambda^i \, ,
\ee
and
\be
c_{2i, 2j}=\zeta_{2i+2j-1} \frac{(-1)^{i+j}}{i+j} {2i+2j \choose 2i} \, .
\ee
Then, for the 2-point functions of even operators, the terms with one and two values of $\zeta$ are
\be
\begin{split}
\vev {\Omega_{2i} \Omega_{2j} }= & f(i)f(j) \left( \frac{1}{i+j} -2\sum_{p,q}c_{2p 2q} f(p)f(q) \left(\frac{1}{p(p+1)}+\frac{1}{(p+i)(q+j)}\right) + \right. \\
 & \frac{1}{2} \sum_{p,q} \sum_{l,m} c_{2p,2q} c_{2l,2m} f(p) f(q) f(l) f(m) \left(\frac{8}{(i+p)(q+l)(m+j)}+ \frac{8}{(q+l)m(m+1)}   
\right. \\
&\left. \left. +  \frac{8}{(l+j)(p+1)p} + \frac{8}{(m+i)(p+1)p}+4 \frac{(i+j+q+m-1)}{(p+1)p(l+1)l}\right) +\dots \right)\, .
\end{split} 
\label{s4twopoint}
\end{equation}
For the 3-point functions, the terms with a single value of $\zeta$ are
\be
\begin{split}
& \langle \Omega_{2\ell_1} \Omega_{2\ell_2} \Omega_{2\ell_3}\rangle = \frac{1}{N} f(\ell_1)f(\ell_2)f(\ell_3) \\
& \left[1-2\sum_{i,j=1}^\infty c_{2i,2j} f(i)f(j) \left(\frac{\ell_1+\ell_2+\ell_3+i-1}{j(j+1)}+\frac{1}{j+\ell_1}+\frac{1}{j+\ell_2}+\frac{1}{j+\ell_3}\right) +\dots \right]\, .
\end{split}
\label{3points4}
\ee
In principle, expanding the determinant (\ref{def:o2k}) allows us to write $O_{2k}$ as linear combinations of $\Omega_{2\ell}$ - see eq. (\ref{o2komega2l}) below - and then we can compute 2- and 3- point functions on $\bR^4$ to the desired order in the number of $\zeta$ values. This is the procedure that we will follow to compute 3-point functions. On the other hand, in the next section we will review how to bypass this approach in the case of 2-point functions.

\section{Two-point functions}
The expression (\ref{def:o2k}) leads to a very compact formula for the two-point function of $O$ operators. Indeed, since the span of $\{\Omega_2,\dots,\Omega_{2k-2}\}$ is the same as the span of $\{O_2,\dots,O_{2k-2}\}$, we have $\langle O_{2k} \bar O_{2k}\rangle = \langle O_{2k} \Omega_{2k}\rangle$. Then using the Laplace expansion of the determinants for the last row, we arrive at
\be
\langle O_{2k} \bar O_{2k} \rangle= \frac{D_k}{D_{k-1}} \, .
\label{ratiogsdet}
\ee
This expression is exact. In what follows we will evaluate it up to terms with two values of $\zeta$. The entries in $D_k$ are two-point functions of operators on $S^4$, and are given by (\ref{s4twopoint}) for terms with up to two values of $\zeta$. The first term
in (\ref{s4twopoint}) corresponds to the ${\cal N}=4$ SYM contribution while the rest arise from the $1-$loop contribution of the $\mathcal{N}=2$ theory \cite{Fiol:2021icm}. It is convenient to give names to the matrices that appear in the various terms of (\ref{s4twopoint}). Let's define the following three $k\times k$ matrices $A,B,C$,

\be
A_{ij}=\frac{1}{i+j} \, ,
\ee
and 
\be
B=B_{ind}+B_{\text{dep}}
=  -2\sum_{p,q}c_{2p 2q} f(p)f(q) \left(\frac{1}{p(p+1)}+\frac{1}{(p+i)(q+j)}\right) \, ,
\label{thetwobs}
\ee
where $B_{\text{ind}}$ denotes the first term, independent of $i,j$, while $B_{\text{dep}}$ refers to the second term, that depends on the indices $i,j$. Finally, 
\begin{equation}
\begin{split}
C_{ij} = & \frac{1}{2} \sum_{p,q} \sum_{l,m} c_{2p,2q} c_{2l,2m} f(p) f(q) f(l) f(m) \left(\frac{8}{(i+p)(q+l)(m+j)}+ \frac{8}{(q+l)m(m+1)}   
\right. \\
&\left. +  \frac{8}{(l+j)(p+1)p} + \frac{8}{(m+i)(p+1)p}+4 \frac{(i+j+q+m-1)}{(p+1)p(l+1)l}\right) \, .
\end{split} 
\end{equation}
The determinant (\ref{gsdet}) of two-point functions is
\be
\begin{split}
D_k= & \left(\prod_{i=1}^k f(i) \right)^2 |A_k| |\bI+A_k^{-1} B+A_k^{-1}C| \\
  =  &
\left(\prod_{i=1}^k f(i) \right)^2 |A_k|\left( 1+\text{tr } (A_k^{-1} B)+ \text{tr } (A_k^{-1} C)+
\frac{1}{2}(\text{tr } (A_k^{-1} B))^2-\frac{1}{2}\text{tr } (A_k^{-1} B)^2 + \dots \right) \, . 
\end{split}
\ee
so according to (\ref{ratiogsdet}), the 2-point functions on $\bR^4$ are then
\be
\begin{split}
& \langle O_{2k} \bar O_{2k}\rangle = \frac{D_k}{D_{k-1}}=f(k)^2 \frac{|A_k|}{|A_{k-1}|} \frac{|\bI+A_k^{-1}B+A_k^{-1}C+\dots |}{|\bI+A_{k-1}^{-1}B+A_{k-1}^{-1}C+\dots |}=\\
& f(k)^2 \frac{|A_k|}{|A_{k-1}|}\left(1+\text{tr }(\Delta_- \cdot B)+\text{tr }(\Delta_- \cdot C)+\frac{1}{2}(\text{tr }(\Delta_- \cdot B))^2-\frac{1}{2}\text{tr } \Delta_+ B\Delta_- B +\dots \right)
\end{split}
\label{twopointtotal}
\ee
where $\Delta_\pm \equiv A_k^{-1} \pm A_{k-1}^{-1}$ and we have used
\be
\text{tr }(A_k^{-1}B)^2-\text{tr } (A_{k-1}^{-1} B)^2=\text{tr } (A_k^{-1}+A_{k-1}^{-1})B(A_k^{-1}-A_{k-1}^{-1})B \, .
\ee
The crucial observation that will allow us to proceed is that the matrix A is of the Cauchy type, so using the general formulas in appendix \ref{cauchy}, we can find its determinant
\be
|A_k|= \frac{\left(\prod_{i=1}^k i!\right)^3}{k! \prod_{i=0}^k (k+i)!} \, ,
\label{detofa}
\ee
and its inverse,
\be
\left(A^{-1}_k \right)_{ij} =\frac{(-1)^{i+j}}{i+j} \frac{(k+i)! (k+j)!}{(k-i)! (k-j)! i! j! (i-1)! (j-1)!} \, .
\ee
As a first application, making use of (\ref{detofa}) the prefactor of (\ref{twopointtotal}) reduces to
\be
f(k)^2 \frac{|A_k|}{|A_{k-1}|}=2k \tilde \lambda^{2k} \, .
\label{prefactor}
\ee
\subsection{Two-point functions up to order $\zeta$}
Let's start by proving the conjectured formula \cite{Fiol:2021icm} for the terms with a single value of $\zeta$ in the two-point function. Keeping terms with a single $\zeta$ in (\ref{twopointtotal}),
\be
\langle O_{2k} \bar O_{2k}\rangle=\frac{D_k}{D_{k-1}}= f(k)^2 \frac{|A_k|}{|A_{k-1}|} \left(1+\text {tr } \Delta_-\, B +\dots \right) \, .
\label{twopointfirst}
\ee
 the trace in (\ref{twopointfirst}) simplifies to
\be
\text {tr } \Delta_- \, B=
2k\sum_{i,j=1}^k (-1)^{i+j} {k \choose i}{k+i-1 \choose i-1}{k \choose j}{k+j-1 \choose j-1} B_{ij} \, .
\label{tracediff}
\ee
It is convenient to split this sum into two terms, for $B_{ind}$ and $B_{dep}$, defined in (\ref{thetwobs}),
\be
\text {tr } \Delta_- \, B=\text {tr } \Delta_- \, B_{dep}+\text {tr } \Delta_- \, B_{ind}
\ee
Let's start computing tr $\Delta_- B_{ind}$. 
In this case, since $B_{ind}$ is independent of $i,j$, we just have to perform the sum over $i,j$, and then multiply by the contribution from $B_{ind}$. The sum over $i,j$ in (\ref{tracediff}) is immediate to carry out, using (\ref{sum1}) twice
\be
2k \sum_{i=1}^k (-1)^i {k \choose i}{k+i-1 \choose k} \sum_{j=1}^k (-1)^j {k \choose j}{k+j-1 \choose k}
=2k (-1)^{2k}=2k \, .
\ee
We now proceed to evaluate $B_{ind}$,
\be
\begin{split}
& B_{ind}=-\sum_{p,q=1}^\infty \frac{\zeta_{2p+2q-1}(-1)^{p+q}}{p+q} {2p+2q \choose 2p} \frac{(2p)! (2q)! \tilde \lambda^{p+q}}{p! (p-1)! q! (q-1)!} \left[\frac{1}{p(p+1)}+\frac{1}{q(q+1)}\right]= \\
&-2\sum_{n=2}^\infty \frac{\zeta_{2n-1} (-\tilde \lambda)^n}{n} {2n \choose n} \sum_{q=1}^{n-1} {n \choose q}{n \choose q-1}=
-2\sum_{n=2}^\infty \frac{\zeta_{2n-1} (-\tilde \lambda)^n}{n} {2n \choose n} \left[ {2n \choose n+1}-n\right]
\end{split}
\label{bindsum}
\ee
where we have defined $n=p+q$ and used (\ref{sum5}) in the last step. Thus,
\be 
\text{tr }\Delta_- B_{ind}=-4k \sum_{n=2}^\infty \frac{\zeta_{2n-1} (-\tilde \lambda)^n}{n} {2n \choose n} \left[ {2n \choose n+1}-n\right]
\ee
Let's now compute tr $\Delta_- B_{dep}$. 
\be 
\text{tr }\Delta_- B_{dep}=-4k \sum_{p,q=1}^\infty c_{2p,2q} f(p)f(q) \sum_{i=1}^k \frac{(-1)^i}{p+i} {k \choose i} {k+i-1 \choose k} \sum_{j=1}^k \frac{(-1)^j}{q+j} {k \choose j} {k+j-1 \choose k}
\ee
Using (\ref{sum4}) twice, this trace reduces to

\be
\begin{split}
& \text{tr }\Delta_- B_{dep}=-4k \sum_{p,q=1}^\infty \frac{\zeta_{2p+2q-1}(-\tilde \lambda)^{p+q}}{p+q} \frac{(2p+2q)!}{(p+k)! (p-k)! (q+k)! (q-k)!}
=\\
& -4k \sum_{n=2}^\infty \frac{\zeta_{2n-1} (-\tilde \lambda)^n}{n} {2n \choose n}\sum_{q=1}^{n-1} {n \choose q+k} {n \choose q-k}
= -4k \sum_{n=2}^\infty \frac{\zeta_{2n-1} (-\tilde \lambda)^n}{n}
{2n \choose n}{2n \choose n+2k}
\label{bdepsum}
\end{split}
\ee
where in the last step we used (\ref{sum6}). Finally, plugging (\ref{prefactor}),(\ref{bindsum}) and (\ref{bdepsum})  in (\ref{twopointfirst}) we arrive at
\be
\langle O_{2k} \bar O_{2k}\rangle = 2k \tilde \lambda^{2k} \left(1-4k\sum_{n=2}^\infty \frac{\zeta_{2n-1} (-\tilde \lambda)^n}{n}
{2n \choose n} \left[ {2n \choose n+2k}+{2n \choose n+1}-n \right] + \mathcal{O}(\zeta^2)\right)
\label{twopointonezeta}
\ee
This concludes the proof of the formula for the term swith one value of $\zeta$ in the planar two-point functions $\langle O_{2k} \bar O_{2k}\rangle$. 

\subsection{Two-point functions up to order $\zeta^2$}
We can now extend the computation to include the $\zeta^2$ terms. There are three contributions at order $\zeta^2$ in (\ref{twopointtotal}). The first one, $(\text{tr }(\Delta  B))^2$, is just the square of the terms with one $\zeta$, that we have already computed in (\ref{twopointonezeta}),
\be
\begin{split}
\frac{1}{2} (\text{tr }(\Delta  B))^2= \frac{1}{2}(-4k)^2  \sum_{m=2}^\infty \frac{\zeta_{2m-1} (-\tilde \lambda)^m}{m}
{2m \choose m} \left[ {2m \choose m+2k}+{2m \choose m+1}-m \right]  \\
\sum_{n=2}^\infty \frac{\zeta_{2n-1} (-\tilde \lambda)^n}{n}
{2n \choose n} \left[ {2n \choose n+2k}+{2n \choose n+1}-n \right]
\end{split}
\ee
The second contribution is $\text{tr } \Delta_- C$. A rather long computation yields 
\be
\begin{split}
& \text{tr } \Delta_- C=  k \sum_{m,n=2}^\infty \frac{\zeta_{2m-1} \zeta_{2n-1} (-\tilde \lambda)^{m+n}}{m n}  \left[16 {2n \choose n} {2m \choose m+k} n ({\cal C}_n-1)(m+k) {2m-1 \choose m+k}  \right. \\
& +8 {2n \choose n+k}{2m \choose m+k} \sum_{i=1}^{n-1} \sum_{j=1}^{m-1} {n+k \choose i}{n-k \choose i}{m+k \choose j}{m-k \choose j} \frac{ij}{i+j} \\
& +8 {2n \choose n-1}{2m \choose m} \sum_{i=1}^{n-1} \sum_{j=1}^{m-1} {n-1 \choose i}{n-1 \choose i-1}{m \choose j}{m \choose j-1} \frac{1}{i+j} \\
& +4 {2n \choose n} {2m \choose m} m n \left[ (2k^2-1) ({\cal C}_m-1)({\cal C}_n-1) +({\cal C}_n-1) \left({2m-1 \choose m}-m\right)  \right.\\
&\left . \left. +({\cal C}_m-1)\left({2n-1 \choose n}-m\right) \right] \right]\\
\end{split}
\ee
Finally, the last contribution is $\text{tr } \Delta_+ B\Delta_-  B$. We split this last computation in terms of $B^{ind}$, and $B^{dep}$. We find
\be
\begin{split}
&\text{tr }\left(\Delta_+  B^{ind}  \Delta_-  B^{ind}\right)= \\
&16 k^3 \sum_{m,n=2}^\infty \frac{\zeta_{2m-1} \zeta_{2n-1} (-\tilde \lambda)^{m+n}}{m\, n} 
{2m \choose m} {2n \choose n} \left[{2m \choose m+1}-m\right]  \left[{2n \choose n+1}-n\right] 
\end{split}
\ee

\be
\begin{split}
&\text{tr }\left(\Delta_+  B^{dep}  \Delta_-  B^{ind}+\Delta_+  B^{ind} \Delta_-  B^{dep}\right)= 
32k \sum_{m,n=2}^\infty \frac{\zeta_{2m-1} \zeta_{2n-1} (-\tilde \lambda)^{m+n}}{m\, n} \\
&{2m \choose m}  \left[{2m \choose m+1}-m\right]  \left[(n+k){2n \choose n+k}{2n-1 \choose n+k}-\frac{n}{2}{2n \choose n}{2n \choose n+2k}\right] 
\end{split}
\ee

\be
\begin{split}
&\text{tr }\left(\Delta_+  B^{dep}  \Delta_-  B^{dep}\right)= 
16 k \sum_{m,n=2}^\infty \frac{\zeta_{2m-1} \zeta_{2n-1} (-\tilde \lambda)^{m+n}}{m n} \\
& \left[ {2n \choose n+k}{2m \choose m+k} \sum_{i=1}^{n-1} \sum_{j=1}^{m-1} {n+k \choose i}{n-k \choose i}{m+k \choose j}{m-k \choose j} \frac{ij}{i+j} \right. \\
& \left. -{2n \choose n}{2m \choose m} \sum_{i=1}^{n-1} \sum_{j=1}^{m-1} {n \choose i+k}{n \choose i-k}{m \choose j+k}{m \choose j-k} \frac{k^2+ij}{i+j} \right]
\end{split}
\ee
Putting these three contributions together, after some simplifications we arrive at 
\be
\begin{split}
&\langle O_{2k}\bar O_{2k} \rangle = 2k \tilde \lambda^{2k} \left[1+{\cal O}(\zeta)+4k \sum_{m,n=2}^\infty \frac{\zeta_{2m-1} \zeta_{2n-1}}{mn}(-\tilde \lambda)^{m+n} {2m \choose m}{2n\choose n} \bigg(  m\, n \bigg[-({\cal C}_m-1)({\cal C}_n-1)\right.  \\
& \left. +({\cal C}_m-1) \left( {2n \choose n+2k}+{2n-1 \choose n}-n\right)+
 ({\cal C}_n-1) \left( {2m \choose m+2k}+{2m-1 \choose m}-m\right)\right] \\
& +2k\left[{2m \choose m+2k}+m({\cal C}_m-1)\right]\left[{2n \choose n+2k}+n({\cal C}_n-1)\right] \\
& +2n^2 \sum_{i=1}^{n-1}{n-1 \choose i}{n-1 \choose i-1} \sum_{j=1}^{m-1}{m\choose j}{m \choose j-1} \frac{1}{i+j} \\
& \left. \left. +2k \sum_{i=1}^{n-1}{n \choose i+k}{n \choose i-k} \sum_{j=1}^{m-1}{m\choose j+k}{m \choose j-k} \frac{k^2+ij}{i+j} \right)+\dots \right]
\end{split}
\label{twozetas}
\ee
Despite many efforts, we haven't been able to simplify the double sums in the last two lines of (\ref{twozetas}). A rather non-trivial check of the validity of (\ref{twozetas}) is that it reproduces all the terms with two values of $\zeta$ for the expressions of $\langle O_2 \bar O_2\rangle$, $\langle O_4 \bar O_4\rangle$, $\langle O_6 \bar O_6\rangle$ available in \cite{Galvagno:2020cgq}.

\section{Three-point functions}
The computation of the terms with a single value of $\zeta$ in three-point functions is substantially more complicated than the equivalent computation for the two-point functions, since it can't be reduced to the evaluation of Gram-Schmidt determinants. The strategy that we will follow is to expand by its last row the determinant relating $O_{2k}$ to $\Omega_{2\ell}$, eq. (\ref{def:o2k}). This expansion involves minors of that matrix, and that makes the expressions more complicated. Consider the following $(k-1)\times (k-1)$ minor,
\be
M_\ell=
\begin{pmatrix}
\frac{1}{1+1} & \frac{1}{2+1}&\dots& \cancel{\frac{1}{\ell+1}} & \dots &\frac{1}{k+1} \\
\frac{1}{1+2} & \frac{1}{2+2}&\dots& \cancel{\frac{1}{\ell+2}} & \dots &\frac{1}{k+2} \\
\dots & \dots & \dots & \dots & \dots & \dots \\
\frac{1}{1+k-1} & \frac{1}{2+k-1}&\dots& \cancel{\frac{1}{\ell+k-1}} & \dots &\frac{1}{k+k-1} 
\end{pmatrix}
\, .
\ee
In terms of this matrix we have
\be   
O_{2k}= \sum_{\ell=1}^k \frac{|M_\ell|}{D_{k-1}} \left[1+\text{tr }M_\ell^{-1}B-\text{tr }A_{k-1}^{-1}B+\dots\right]\Omega_{2\ell} \, .
\label{ofromomega}
\ee
The minor M is still a Cauchy matrix, so using the general formulas of the appendix \ref{cauchy}, we can compute its determinant,
\be
\det M_\ell =
\frac{(\ell+k-1)! k!}{\ell ! (2k-1)! (k-\ell)! (\ell-1)!} \frac{\left( \prod_{i=1}^{k-1} i!\right)^3}{\prod_{i=0}^{k-1} (k-1+i)!} \, ,
\ee
and its inverse
\be
\left(M^{-1}_\ell\right)_{ij}=\frac{ (-1)^{\bar i+j} (k+\bar i-1)! (j+k)! (\ell-\bar i)}{(\bar i+j) \bar i! (\bar i-1)! j! (j-1)! (k-\bar i)! (k-1-j)! (j+\ell)} \, ,
\ee
where $\bar i$ is defined by
\be
\bar i=
\left\{
\begin{matrix}
i & i < \ell \\
i+1 & i \geq \ell
\end{matrix}
\right.
\ee
As a check, $M_k^{-1}=A^{-1}_{k-1}$ and $|M_k|=|A_{k-1}|$. We can now write the relation between $O_{2k}$ and $\Omega_{2\ell}$ operators, at order $\zeta$,
\be
O_{2k}=2k \sum_{\ell=1}^{k} (-1)^{k+\ell} \tilde \lambda^{k-l} \frac{(\ell+k-1)!}{(2\ell)! (k-\ell)!} 
\left(1+\text{tr }N(k,\ell)\cdot B+\dots \right) \Omega_{2\ell} \, ,
\label{o2komega2l}
\ee
where 
\be
N(k,\ell)=(-1)^{i+j} {k+i-1 \choose k} {k \choose i} {k+j-1 \choose k} {k \choose j}\frac{\ell-k}{\ell+j} \, .
\ee
A non-trivial check of (\ref{o2komega2l}) is that it reproduces the linear terms in $\zeta$ of $\langle O_{2k}\bar O_{2k}\rangle$. We have now all the ingredients to compute the three-point function of $O_{2k}$ operators to order $\zeta$. 
It is actually easier to compute the normalized 3-point functions. Define
\be
P_{ij}(k,\ell)=(-1)^{i+j} {k \choose i}{k+i-1\choose k}{k \choose j}{k+j-1\choose k} \frac{j^2-k^2-j(\ell+j)}{\ell+j} \, .
\label{thep}
\ee

Then,
\be
\begin{split}
& \langle O_{2k_1}O_{2k_2}\bar O_{2k_3}\rangle_n = (2k_1 2k_2 2k_3)^{1/2} \sum_{\ell_1=1}^{k_1}
(-1)^{\ell_1} {k_1+\ell_1-1 \choose k_1}{k_1 \choose \ell_1 } \sum_{\ell_2=1}^{k_2} (-1)^{\ell_2} {k_2+\ell_2-1 \choose k_2}{k_2 \choose \ell_2 }
 \\
& \sum_{\ell_3=1}^{k_3} (-1)^{\ell_3} {k_3+\ell_3-1 \choose k_3}{k_3 \choose \ell_3 }
\Bigg(1+\text{tr } P(k_1,\ell_1)B+\text{tr } P(k_2,\ell_2)B+ \text{tr } P(k_3,\ell_3)B  \\
& \left. - 2\sum_{p,q=1}^\infty c_{2p,2q} f(p)f(q) \left(\frac{\ell_1+\ell_2+\ell_3+p-1}{q(q+1)}+\frac{1}{p+\ell_1}+\frac{1}{p+\ell_2}+\frac{1}{p+\ell_3}\right) \right) \, .
\end{split}
\label{norm3point}
\ee
Actually, the $j^2-k^2$ term in (\ref{thep}) vanishes after summing over $\ell_i$ , so under the sum signs we are allowed to just keep
\be
P_{ij}(k )=-j (-1)^{i+j} {k \choose i}{k+i-1\choose k}{k \choose j}{k+j-1\choose k}  \, .
\ee
There are two sources of ${\cal O}(\zeta)$ terms in $\langle O_{2k_1} O_{2k_2} \bar O_{2k_1+2k_2}\rangle$. The first comes from the order ${\cal O}(\zeta)$ in $\langle \Omega_{2\ell_1} \Omega_{2\ell_2} \Omega_{2\ell_3}\rangle$ in eq. (\ref{3points4}). The second source is the ${\cal O}(\zeta)$ contribution to the coefficient of $\Omega_{2\ell}$ in $O_{2k}$ in the equation (\ref{o2komega2l}) above.

Let's start with the $\ell_i$-independent term in (\ref{norm3point}). Taking into account (\ref{sum5a}) we obtain
\be
\begin{split}
&-2\sum_{p,q}c_{2p,2q} f(p)f(q) \frac{p-1}{q(q+1)}=-2\sum_{p,q} \zeta_{2p+2q-1} \frac{(-\tilde \lambda)^{p+q}}{p+q} \frac{(2p+2q)!}{p!(p-2)!q!(q+1)!}=\\
&-2 \sum_{n=2}^\infty \zeta_{2n-1} \frac{(-\tilde \lambda)^n}{n} \frac{(2n)!}{n! (n-1)!} \sum_{q=1}^{n-1} {n \choose q}{n-1 \choose q+1}
=-\sum_{n=2}^\infty \zeta_{2n-1} (-\tilde \lambda)^n {2n \choose n} (n-1)\left({\cal C}_n-2\right)
\end{split}
\label{lind3point}
\ee
The $\ell_i$-dependent terms in (\ref{norm3point}) are three copies of a common pattern, so we can focus on just one of these three copies. It has four contributions, and it is convenient to pair them as follows. First, pair these two terms
\be
\sum_{\ell=1}^k (-1)^\ell {k+\ell-1 \choose k}{k \choose \ell} \left(\text{tr }P(k,\ell)\cdot B_{ind}-2\sum_{p,q} c_{2p,2q} f(p)f(q)\frac{\ell}{q(q+1)}\right)=0
\ee
That they add up to zero is seen by explicitly plugging in $B_{ind}$. The only remaining terms (one for each $k_i$) are of the form 
\be
\begin{split}
&(-1)^k \sum_{\ell=1}^k (-1)^\ell {k+\ell-1 \choose k} {k \choose \ell} \left[-2 \sum_{p,q} c_{2p,2q} f(p) f(q) \right. \\
&\left. \left(\frac{1}{p+\ell}+\sum_{i,j=1}^k (-1)^{i+j} {k+i-1\choose k} {k \choose i}{k+j-1\choose k} {k \choose j} \frac{1}{p+i} \frac{-j}{q+j} \right) \right] 
\end{split}
\ee
To proceed, exchange the order of sums, use (\ref{sum1}) and (\ref{sum4}) twice so the previous expression simplifies to
\be
-\sum_{n=2}^\infty \zeta_{2n-1} (-\bar \lambda)^n {2n \choose n} \sum_{q=1}^{n-1} {n \choose q+k}{n \choose q-k}q=
-\sum_{n=2}^\infty \zeta_{2n-1} (-\bar \lambda)^n {2n \choose n} {2n \choose n+2k} \, ,
\label{ldep3point}
\ee
where in the last step we have used (\ref{sum7}). Plugging the partial results (\ref{lind3point}) and (\ref{ldep3point}) into (\ref{norm3point}) we finally arrrive at
\be
\begin{split}
&\langle O_{2k_1}O_{2k_2}\bar O_{2k_3}\rangle_n = (2k_1 2k_2 2k_3)^{1/2}\\
&\left [1-\sum_{n=2}^\infty \zeta_{2n-1} (-\tilde \lambda)^n {2n \choose n}
\left ({2n \choose n+2k_1}+{2n \choose n+2k_2}+{2n \choose n+2k_3} +(n-1)\left({\cal C}_n-2\right)\right ) +\dots \right]
\end{split}
\ee
proving the conjecture of \cite{Fiol:2021icm} for the planar limit of three-point function of extremal correlators.

\acknowledgments
Research supported by  the State Agency for Research of the Spanish Ministry of Science and Innovation through the ``Unit of Excellence Mar\'ia de Maeztu 2020-2023'' award to the Institute of Cosmos Sciences (CEX2019-000918-M) and PID2019-105614GB-C22, and by AGAUR, grant 2017-SGR 754.  A. R. F. is further supported by an FPI-MINECO fellowship. 

 \appendix

\section{Cauchy matrices}
\label{cauchy}

By definition, a Cauchy matrix A is a square matrix whose entries $a_{ij}$ are of the form
$$
a_{ij}=\frac{1}{x_i-y_j} \, ,\hspace{1cm} x_i-y_j\neq 0 \, ,
$$
for some sequences $x_i,y_j$ such that $x_i \neq x_j$ and $y_i\neq y_j$.

The determinant of a Cauchy matrix is \cite{schechter}
$$
|A|= \frac{\prod_{i=2}^n \prod_{j=1}^{i-1} (x_i-x_j)(y_j-y_i)}{\prod_{i=1}^n \prod_{j=1}^n (x_i-y_j)} \, .
$$
The determinant is always non-zero, so Cauchy matrices are always invertible. To display the elements of the inverse of a Cauchy matrix, first define
$$
C(x)=\prod_{i=1}^n (x-x_i) \, , \hspace{1cm} D(x)=\prod_{i=1}^n (x-y_i) \, ,
$$
and
$$
C_i(x)=\frac{C(x)}{C'(x_i)(x-x_i)} \, , \hspace{1cm} B_i(x)=\frac{B(x)}{B'(y_i)(x-y_i)} \, .
$$
Then
$$
(A^{-1})_{ij}= (x_j-y_i) C_j(y_i) D_i(x_j) \, .
$$

\section{Some sums}
In this appendix we collect various sum identities that we use in the main text. They are grouped according to the techniques we have used to prove them, and we provide a proof of one of the identities in each group. An excellent source of techniques to deal with sums of binomial coefficients is chapter 5 of \cite{concrete}.

\subsection{A first family of identities}
The combinatorial Vandermonde identity
\be
\sum_{p=0}^r {n \choose p}{m \choose r-p} = {n+m \choose r} \, ,
\ee
can be proven by comparing the coefficients of $x^r$ in
\be
(1+x)^{n+m}=(1+x)^n (1+x)^m =\sum_{p=0}^n {n \choose p}x^p \sum_{q=0}^m {m \choose q}x^q \, .
\ee
If we take one derivative of $(1+x)^n$ above, we arrive at the related identity
\be
\sum_{p=0}^r p {n \choose p}{m \choose r-p} = n {n+m-1 \choose r-1} \, .
\ee
The first family of identities that we use in the main text can all be proven from the two identities above. They are
\be
\sum_{q=1}^{n-1} {n \choose q}{n\choose q-1} =n( C_n-1) \, ,
\label{sum5}
\ee

\be
\sum_{q=1}^{n-1} {n \choose q}{n-1\choose q+1} = (n-1)(C_n-2)/2 \, ,
\label{sum5a}
\ee

\be
\sum_{q=1}^{n-1} {n \choose q}{n\choose q-1} q =n \left( {2n-1 \choose n}-n \right) \, ,
\label{sum5b}
\ee

\be
\sum_{q=1}^{n-1} {n+k \choose q}{n-k \choose q} = {2n \choose n+k}-1 \hspace{1cm} n\ge  k \, ,
\label{sum5c}
\ee

\be
\sum_{q=1}^{n-1} {n+k \choose q}{n-k \choose q}q  = (n+k) {2n-1 \choose n+k} \hspace{1cm} n\ge  k \, ,
\label{sum5d}
\ee

\be
\sum_{q=1}^{n-1} {n \choose q-k}{n\choose q+k} ={2n \choose n+2k} \, ,
\label{sum6}
\ee

\be
\sum_{q=1}^{n-1} {n \choose q-k}{n\choose q+k} q=\frac{n}{2}{2n \choose n+2k} \, ,
\label{sum7}
\ee

\subsection{A second family of identities}
The second family of identities that we use in the main text can be proven with a similar strategy to the one above, but now involving binomials with negative powers. Namely
\be
\frac{1}{(1+x)^{n+1}}=\sum_{k=0}^\infty {n+k \choose n} (-x)^k \, ,
\ee
and
\be
(1+x)^n =\sum_{k=0}^n {n\choose k} x^k \, ,
\ee
For instance
\be
\sum_{\ell=1}^k (-1)^\ell {k+\ell-1 \choose k}{k \choose \ell}= \frac{-x (1+\frac{1}{x})^k}{(1+x)^{k+1}} \vert_{x^0}=\frac{-x^{1-k}}{1+x}\vert_{x^0}=(-1)^k \, .
\label{sum1}
\ee

Following similar steps, one can prove the following identities
\be
\sum_{\ell=1}^k (-1)^\ell \frac{(k+\ell-1)!}{\ell! (\ell-1)! (k-\ell-1)!} =(-1)^{k+1}\, k(k-1) \, ,
\label{sum2}
\ee

\be
\sum_{i=1}^k (-1)^i {k+i-1 \choose k} {k \choose i} i =(-1)^k k^2 \, .
\label{sum3}
\ee

\subsection{A third family of identities}
The third family of sums we have encountered involve sums over binomial coefficients and fractions. Let's prove the first one
\be
\sum_{i=1}^k (-1)^i {k+i-1 \choose i-1} {k \choose i} \frac{1}{i+j} = (-1)^k \frac{j!}{(j+k)!} \frac{(j-1)!}{(j-k)!} \, .
\label{sum4}
\ee 
To prove it, define
\be
g(k,p)=\sum_{i=1}^k (-1)^i {k \choose i}{k+i-1 \choose k} \frac{1}{i+p} \, .
\ee
To evaluate $g(k,p)$, notice that $g(1,p)=-1/(1+p)$ and that it satisfies the recursion relation
\be
g(k+1,p)=-\frac{p-k}{p+k+1}g(k,p) \, ,
\ee
it follows that 
\be
g(k,p)=(-1)^k \frac{p!}{(p+k)!}\frac{(p-1)!}{(p-k)!} \, .
\ee
proving the claim. An easy consequence of (\ref{sum4}) is that
\be
\sum_{i=1}^k (-1)^i {k+i-1 \choose i-1} {k \choose i} \frac{i}{i+j} = (-1)^k \left(1-\frac{j!}{(j+k)!} \frac{j!}{(j-k)!}\right) \, .
\label{sum4b}
\ee

\bibliographystyle{JHEP}

\end{document}